\pdfoutput=1
\documentclass[aps,prl,english,twocolumn,noshowpacs,a4paper,superscriptaddress]{revtex4-1}

\usepackage{graphicx}
\usepackage{color}
\usepackage{textcomp}
\usepackage{amssymb}\usepackage{amsmath}\usepackage{amsthm}
\usepackage{units} 
\usepackage[british,english,ngerman]{babel}
\usepackage{setspace}

\newcommand{\ahat}{\hat{a}}
\newcommand{\adag}{\hat{a}^\dag}
\newcommand{\bhat}{\hat{b}}
\newcommand{\bdag}{\hat{b}^\dag}
\newcommand{\chat}{\hat{c}}
\newcommand{\cdag}{\hat{c}^\dag}
\newcommand{\dhat}{\hat{d}}
\newcommand{\dddag}{\hat{d}^\dag}
\newcommand{\ain}{\hat{a}_\mathrm{in}}

\newcommand{\cin}{\hat{c}_\mathrm{in}}

\newcommand{\bra}{\langle}
\newcommand{\ket}{\rangle}

\begin{document}
\selectlanguage{english} 
\title{Measuring the dynamic structure factor of a quantum gas undergoing a structural phase transition }

\author{Renate Landig}
\affiliation{Institute for Quantum Electronics, ETH Z\"{u}rich, CH--8093 Z\"{u}rich, Switzerland}

\author{Ferdinand Brennecke}
\affiliation{Physikalisches Institut, University of Bonn, Wegelerstrasse 8, 53115 Bonn, Germany}

\author{Rafael Mottl}
\affiliation{Institute for Quantum Electronics, ETH Z\"{u}rich, CH--8093 Z\"{u}rich, Switzerland}

\author{Tobias Donner$^*$}
\affiliation{Institute for Quantum Electronics, ETH Z\"{u}rich, CH--8093 Z\"{u}rich, Switzerland}
\email{Email: donner@phys.ethz.ch}

\author{Tilman Esslinger}
\affiliation{Institute for Quantum Electronics, ETH Z\"{u}rich, CH--8093 Z\"{u}rich, Switzerland}

\date{\today}
\begin{abstract}
The dynamic structure factor is a central quantity describing the physics of quantum many-body systems, capturing  structure and collective excitations of a material. In condensed matter, it can be measured via inelastic neutron scattering, which is an energy-resolving probe for the density fluctuations. In ultracold atoms, a similar approach could so far not be applied due to the diluteness of the system. Here, we report on a direct, real-time and non-destructive measurement of the dynamic structure factor of a quantum gas exhibiting cavity-mediated long-range interactions. The technique relies on inelastic scattering of photons, stimulated by the enhanced vacuum field inside a high finesse optical cavity. We extract the density fluctuations, their energy and lifetime while the system undergoes a structural phase transition. We observe an occupation of the relevant quasi-particle mode on the level of a few excitations, and provide a theoretical description of this dissipative quantum many-body system. 

\end{abstract}

\maketitle
An interacting quantum many-body system can be characterized by analyzing its response to a weak perturbation. In the framework of linear response theory a key quantity is the dynamic structure factor, which is the Fourier transform of the spatial and temporal density-density correlations \cite{Hove1954, Pines1999}. Knowledge of the dynamic structure factor provides a complete picture of the emerging quasi-particle modes \cite{Sachdev2000}, their excitation energy, life-time and mean occupation number. These quasi-particle modes determine the collective density fluctuations of the system and may also characterize the critical behavior in the vicinity of a phase transition \cite{Roth2004}. For example, in a long-range interacting system a structural phase transition can be driven by a roton-like mode softening \cite{Santos2003, ODell2003, Henkel2010, Mottl2012a}, which is expected to show up as a thermally enhanced peak in the dynamic structure factor \cite{Klawunn2011}. 

\begin{figure}
 \includegraphics[width=8cm]{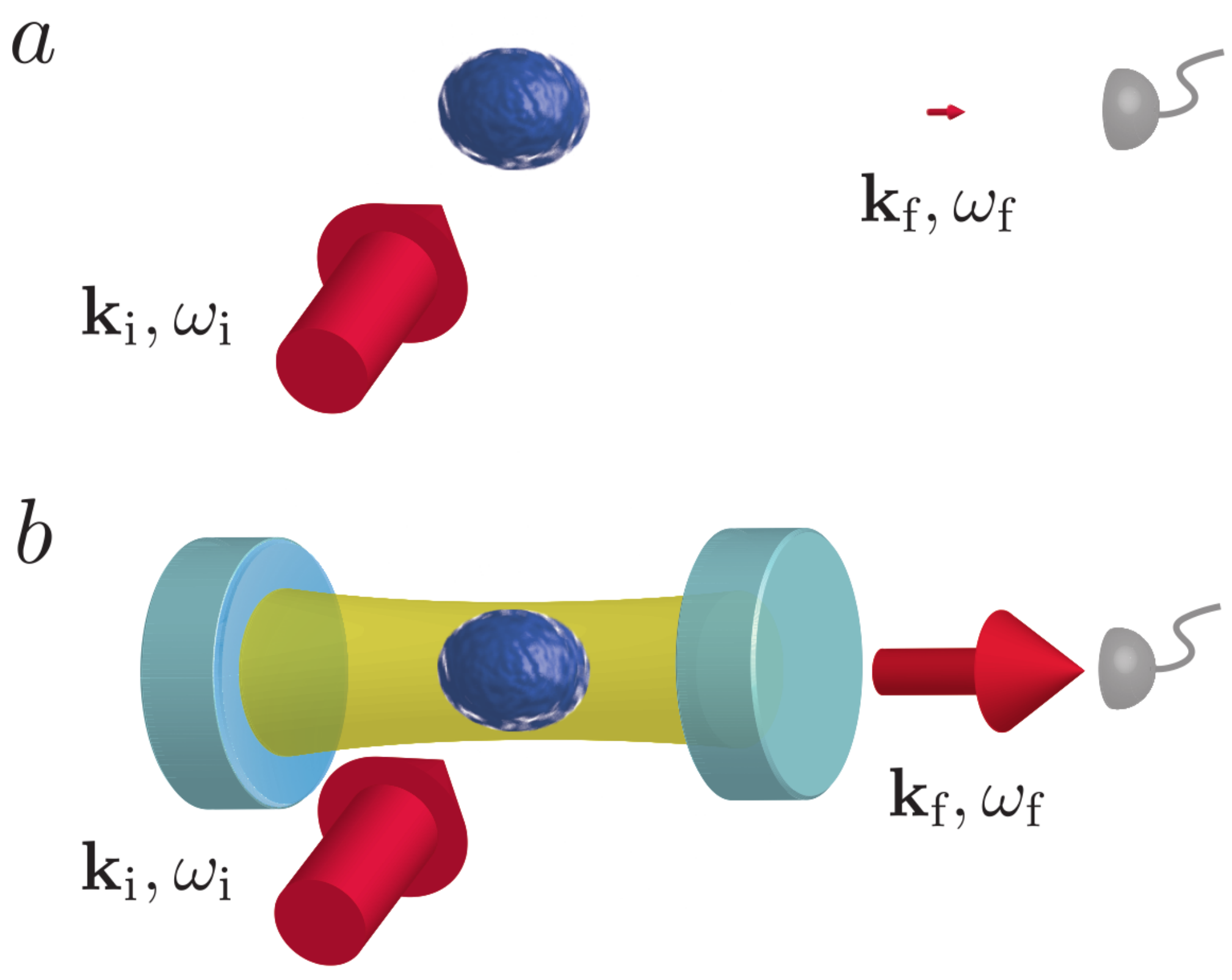}
\caption{\textbf{Scheme for measuring the dynamic structure factor in a quantum gas.} (a) An incident laser beam (red) with wave vector $\mathbf{k}_\mathrm{i}$ is spontaneously scattered at an atomic cloud (blue) into a free-space mode with wave vector $\mathbf{k}_\mathrm{f}$. Analyzing the scattered photons as a function of their frequency shift  $\omega_\mathrm{f}-\omega_\mathrm{i}$ and magnitude yields the dynamic structure factor. The resulting signal for dilute quantum gases is vanishingly small, as indicated by the small size of the arrow pointing towards the detector. (b) Atoms placed into an optical high-finesse resonator (light blue) feel a strongly enhanced vacuum field (golden). Their spontaneous scattering rate into the mode $\mathbf{k}_\mathrm{f}$ can hereby be increased by orders of magnitude, resulting in a detectable signal for the dynamic structure factor.}
 \label{fig:scheme}
\end{figure}

In solid state systems, the dynamic structure factor $S(\mathbf{k},\omega)$ can be measured by illuminating a sample with a beam of neutrons \cite{Squires1978} or x-rays \cite{Sette1998}, and analyzing the inelastically scattered particles with regard to their change in energy $\hbar \omega$ and momentum 􏰀$\hbar \mathbf{k}$. In dilute quantum gases, a measurable signal has only been obtained from photons elastically scattered off a density-modulated sample \cite{Weidemuller1995, Birkl1995, Miyake2011, Corcovilos2010}. Direct detection of inelastically scattered photons into free space \cite{Javanainen1995}, in analogy to neutron scattering, is however hindered by a vanishingly small signal \cite{Stamper-Kurn2001a}, see Fig. 1A. 

A technique measuring the spectral response function of a quantum gas, i.e. the dynamic structure factor at zero temperature \cite{Pitaevskii2003}, is Bragg spectroscopy \cite{Stenger1999, Steinhauer2002}. It is based on stimulated rather than spontaneous inelastic scattering of photons between two laser beams. Therefore the transfer of momentum and energy to the atomic cloud is predetermined by the angle and frequency difference between the beams, and is typically measured via destructive absorption imaging. A complementary detection method analyzes the change in light field intensity in one of the two Bragg beams \cite{Pino2011}, and could in principle be extended with the help of cavities to be only weakly perturbative \cite{Weimer2011}. However, all these methods measure the linear response of the gas upon a perturbation and are insensitive to thermally excited quasi-particles. A different approach, \emph{in-situ} imaging, has been used to extract the temperature-dependent static structure factor $S(\mathbf{k}) = \int \mathrm{d}\omega S(\mathbf{k},\omega)􏰅$ of a two-dimensional gas  \cite{Hung2011}. Similar to the analysis of noise correlations from images of ballistically expanded ultracold gases \cite{Folling2005}, this approach gives no access to the quasi-particle spectrum, i.e. the temporal dynamics.

Here, we present a non-destructive, direct measurement of the dynamic structure factor at distinct $\mathbf{k}$-vectors in a Bose-Einstein condensate (BEC) undergoing a structural phase transition induced by cavity-mediated long-range interactions. We place a BEC into an ultra-high finesse optical cavity \cite{Ritsch2013a, Brennecke2007}  and illuminate the atoms with a transverse laser field. The enhanced vacuum field inside the optical resonator \cite{Haroche2013} increases the spontaneous inelastic scattering of photons into the cavity mode by several orders of magnitude, so that photons leaking out of the cavity mode give rise to a detectable signal and access to the density correlations in real time, see Fig. 1B. Therefore density fluctuations in the gas are mapped onto fluctuations of the light field, which then can be directly accessed.
\begin{figure}[h]
 \includegraphics[]{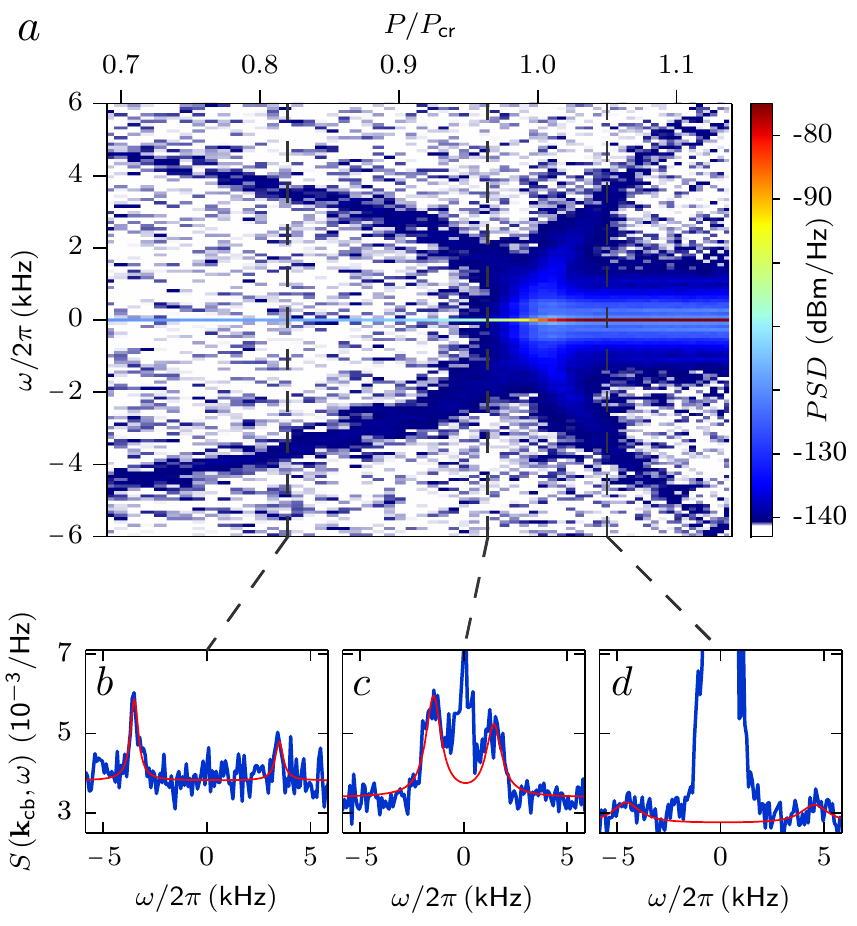}
\caption{\textbf{Power spectral density and dynamic structure factor.} (a) The power spectral density $PSD$ of the light field leaking out of the cavity is shown as a function of frequency shift $\omega$ with respect to $\omega_\mathrm{p}$ and relative transverse pump power $P/P_\mathrm{cr}$ (average over 147 experimental repetitions). Two sidebands are visible, corresponding to the incoherent creation ($\omega<0$) and annihilation ($\omega>0$) of quasi-particles. The energy of these quasi-particles vanishes towards the critical point. At the phase transition, a strong coherent field at the pump frequency appears ($\omega=0$). We attribute the broadened feature around $\omega = 0$ to residual low-frequency technical noise in our system. Note the large dynamic range of the data on the logarithmic scale. The panels (b-d)  show the normalized dynamic structure factor $S(\mathbf{k}_\mathrm{cb},\omega)$ for three different values of $P/P_{\mathrm{cr}}$ (see dashed lines in upper panel), derived from the power spectral density $PSD(\omega)$. While the position and width of the sidebands give direct access to the energy and lifetime of the quasi-particles, the sideband asymmetry can be used to determine the occupation of the quasi-particle mode. Red line shows a fit to the sidebands with our theoretical model to extract these properties.}
\label{fig:spectrum}
\end{figure}

\section*{Results}
\subsection{System description} In our experiment, the transverse laser field acts simultaneously as a pump field controlling the long-range interactions \cite{Baumann2010, Mottl2012a}.  This field has a frequency $\omega_\mathrm{p}$ and a wavevector $\mathbf{k}_\mathrm{p}$, and is in a standing-wave configuration directed perpendicularly to the cavity mode. It is far detuned from atomic resonance to avoid electronic excitation of the atoms. At the same time, it is detuned by only a few cavity linewidths from the cavity resonance, which enables vacuum-stimulated scattering of pump photons into the cavity mode at wavevector $\mathbf{k}_\mathrm{c}$. These two-photon processes mediate long-range atom-atom interactions in the BEC, giving rise to a roton-like mode softening \cite{Mottl2012a} and a structural phase transition \cite{Baumann2010, Brennecke2013a}. The same two-photon processes are exploited for detection. The light scattered from the transverse pump field into the cavity mode can be regarded as a superposition of all field amplitudes scattered by the individual atoms. It thus carries information on the density-density correlations of the gas at the wavevectors $\mathbf{k}_\mathrm{cb}=\pm \mathbf{k}_\mathrm{p} \pm \mathbf{k}_\mathrm{c}$, which are determined by the underlying two-photon processes \cite{Brennecke2013a}.  Within the cavity linewidth, which is two orders of magnitude larger than the frequency of the relevant quasi-particle excitation, the energies of the photons stimulated into the vacuum mode are not fixed. This is in contrast to Bragg spectroscopy, where the energy of the quasi-particle excitations that are created during probing is determined by the frequency difference of the classical fields driving the two-photon processes. The spectral analysis of the light field leaking out of the cavity thus gives us direct access to the dynamic structure factor at finite temperatures.

As previously described \cite{Baumann2010, Baumann2011, Mottl2012a, Brennecke2013a}, we trap a BEC of $N =$ $1.0(1)\times 10^5$ $^{87}$Rb atoms at the center of an ultrahigh-finesse optical Fabry-P\'{e}rot cavity and illuminate it by the transverse pump field. The cavity-mediated interaction leads to the formation of a quasi-particle mode, which is a superposition of the collective momentum excitation at wavevectors $\mathbf{k}_\mathrm{cb}$ of the BEC and a tiny admixture of  photons inside the cavity (see Supplementary Note 1). Neglecting atom-atom collisions, its energy $\hbar \omega_\mathrm{s}$ equals in the limit of zero pump power $P$ the bare energy $\hbar \omega_0 = \hbar^2 k_\mathrm{cb}^2/(2m)$ of a single momentum excitation and decreases with increasing power \cite{Mottl2012a}, where $m$ denotes the atomic mass. Due to this mode softening, the energy $\hbar \omega_\mathrm{s}$ of the quasi-particle mode approaches zero at a critical pump power $P_\mathrm{cr}$, which leads to a phase transition from a normal state with a flat density distribution to a self-organized state with checkerboard density modulation. The emergent density structure leads to elastic scattering of transverse pump light and a macroscopic population of the cavity mode. An order parameter of this phase transition is the expectation value of the operator $\hat{\Theta}$ describing the overlap of the atomic density $\hat{\rho}(\mathbf{r},t)$ and the checkerboard mode structure, $\hat{\Theta}(t) = \int \mathrm{d}^3\mathbf{r} \hat{\rho}(\mathbf{r},t) \cos(\mathbf{k}_\mathrm{c} \mathbf{r})\cos(\mathbf{k}_\mathrm{p} \mathbf{r})$.

\begin{figure}
 \includegraphics[]{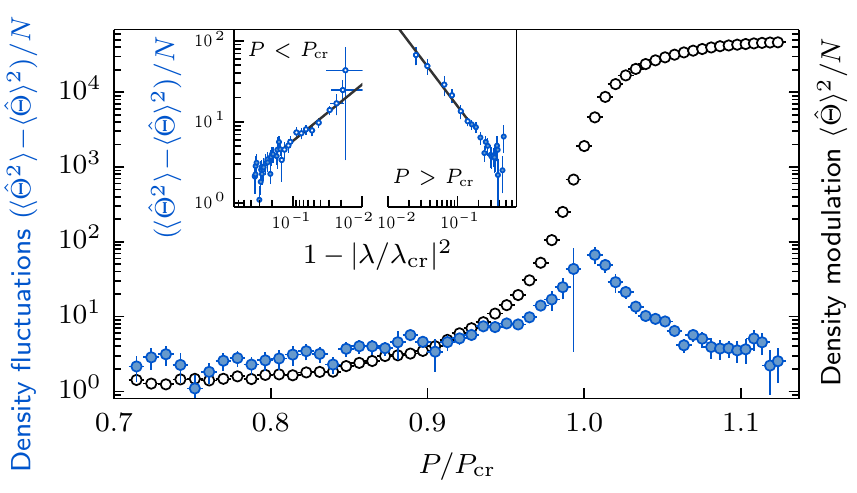}
\caption{\textbf{Density fluctuations and density modulation.} The variance of density fluctuations (filled blue symbols) and the square of the density modulation (open black symbols) of the long-range interacting quantum gas is shown as a function of relative pump power $P/P_\mathrm{cr}$. The fluctuation data is extracted from the dynamic structure factor (see Fig. 2) by integrating over the fit function describing the sidebands and is proportional to the static structure factor. The coherent density modulation is calculated from the power spectral density at the zero frequency bin. The vertical error bars display the statistical error (s. d.) from the fit, while the horizontal error bars display the standard deviation in our determination of the critical point. The inset displays a double logarithmic plot to demonstrate the scaling behavior of the variance of the density fluctuations against the distance to the critical point, expressed as the Hamiltonian coupling parameter $\lambda$ (see Supplementary Note 1). From a linear fit, we find critical exponents of 0.7(1) and 1.1(1) on the normal and self-organized side, respectively. The open symbols in the inset are used for the fitting.}
\label{fig:scaling}
\end{figure}

\subsection{Dynamic structure factor}
As the cavity decay rate $\kappa =2\pi \times \unit[1.25]{MHz}$ is more than two orders of magnitude faster than the evolution rate $\omega_\mathrm{s}$ of the coupled system, the light field $\hat{a}(t)$ inside the cavity adiabatically follows the order parameter, $\hat{a}(t)\propto \hat{\Theta}(t)$ \cite{Brennecke2013a}. The frequency spectrum of the light leaking out of the cavity thus reveals the temporal and spatial Fourier transform of the atomic density correlations, evaluated at one particular wavevector. Specifically, the dynamic structure factor of the system at wavevector $\mathbf{k}_\mathrm{cb}$ is related to the cavity field according to
\begin{equation}\label{eq:structure_factor}
S(\mathbf{k}_\mathrm{cb},\omega) =  \frac{\kappa^2+\tilde{\Delta}_c^2}{\eta^2} \frac{4}{N}  \left( \frac{1}{2\pi} PSD(\omega) - |\alpha|^2 \delta({\omega})\right)\,,
\end{equation}
where $\omega$ is the frequency shift of the cavity output field from the pump light  frequency $\omega_\mathrm{p}$ due to inelastic scattering. $PSD(\omega)$ is the power spectral density of the intra-cavity light field with mean coherent field amplitude $\alpha=\langle \hat{a} \rangle$, and $\eta$ is the two-photon Rabi-frequency of the scattering process, proportional to $\sqrt{P}$. $\tilde{\Delta}_c/2\pi = \unit[15.1(2)]{MHz}$ is the detuning between the pump laser frequency and the dispersively shifted cavity resonance (see Supplementary Note 1). 

To analyze the light field leaking out of the cavity, we use a balanced heterodyne detection scheme \cite{Baumann2011}. Figure 2 shows the power spectral density $PSD(\omega)$ of the light field as we linearly increase the transverse pump power $P$ across the critical point. The rate of change of $P/P_\mathrm{cr}$ is a few Hertz, such that the system can be assumed to adiabatically follow its steady state throughout the measurement \cite{Baumann2011, Brennecke2013a}. The small panels in Fig. 2 show examples of $S(\mathbf{k}_\mathrm{cb},\omega)$ for different values of $P/P_\mathrm{cr}$ , converted via Eq. (\ref{eq:structure_factor}) and normalized to unity for the non-interacting case \cite{Zambelli2000}. The data reflects the microscopic processes taking place: Pump photons of frequency $\omega_\mathrm{p}$ inelastically scattered at the atomic ensemble will be shifted in their frequency. They become visible as red (blue) sideband at frequency $\omega_\mathrm{p}-\omega_\mathrm{s}$ ($\omega_\mathrm{p}+\omega_\mathrm{s}$) if they create (annihilate) a quasi-particle. We observe the corresponding sidebands whose frequency shift tends to zero when approaching the critical point at $P/P_\mathrm{cr}=1$ from either side of the phase transition. These density fluctuations can be distinguished from a checkerboard density modulation at $\mathbf{k}_\mathrm{cb}$ at which pump photons will be elastically scattered without a frequency shift, visible at $\omega=0$. Intuitively this light field arises from scattering at Bragg planes in the density modulated cloud. The transverse pump power $P$ influences not only the effective long-range interactions in the system, but also the measurement process itself. For increasing pump power, the measurement imprecision due to the shot noise of the transverse pump field becomes less relevant, as can be seen from the decreasing background level of $S(\mathbf{k}_\mathrm{cb},\omega)$ \cite{Clerk2010}. The influence of the inevitable measurement backaction will be discussed further below.

The total power of elastically scattered light  is proportional to the square of the density modulation $\bra\hat{\Theta}\ket^2$, and is displayed as open symbols in Fig. 3. This coherent density modulation increases over five orders of magnitude while crossing the critical point. In the normal phase ($P/P_\mathrm{cr}<1$), we also observe a weak field at the pump laser frequency (i.e. at $\omega=0$), which originates from a small symmetry breaking field caused by the finite size of the system and residual scattering of the transverse pump beam at the cavity mirrors \cite{Baumann2011, Brennecke2013a}. 
 
The total power of frequency-shifted light is proportional to the variance of the checkerboard density fluctuation $\bra\hat{\Theta}^2\ket - \bra\hat{\Theta}\ket^2$, and thus to the static structure factor $S(\mathbf{k}_\mathrm{cb}) = \int \mathrm{d}\omega\, S(\mathbf{k}_\mathrm{cb},\omega)$. We show the variance of density fluctuations as filled symbols in Fig. 3, and observe a divergence when approaching the critical point $P/P_\mathrm{cr}=1$ from either side, heralding a second-order phase transition. The inset displays the variance of the density fluctuations on a double logarithmic plot to illustrate the scaling behavior. The variance is plotted as a function of the Hamiltonian coupling parameter $\lambda$, derived from measured quantities and using a theoretical model (see Supplementary Note 1). In the normal and the self-organized phase we extract critical exponents of 0.7(1) and 1.1(1), respectively. We attribute the deviation from our previous measurement in the normal phase, which gave 0.9(1)  \cite{Brennecke2013a}, to the refined model used for the scaling of the horizontal axis and the improved measurement scheme that allows us to directly distinguish between density fluctuations and a density modulation. Current theoretical research taking into account the open character of the system due to cavity dissipation predicts an exponent of 1.0 \cite{Nagy2011, Oztop2012}. The difference between the experimentally observed exponent and the predicted value might originate from finite size effects and the presence of a small symmetry breaking field. Further,  the theory models do not include damping of the momentum excitation due to atom-atom collisions \cite{Kulkarni2013, Konya2014, Konya2014a}.

\begin{figure}
\includegraphics[]{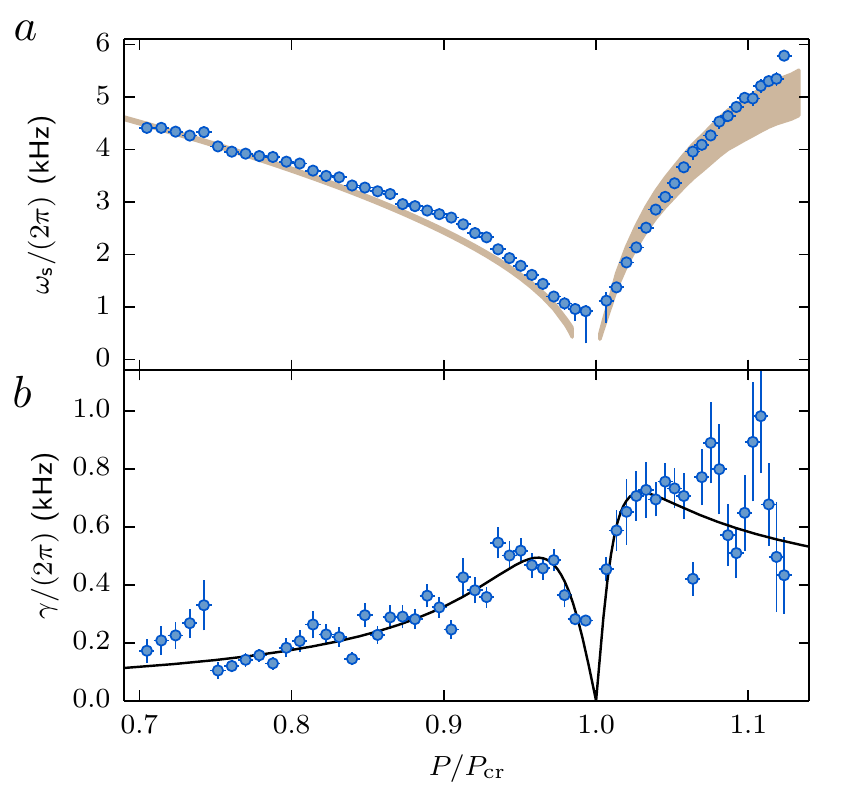}
\caption{\textbf{Characterization of the quasi-particle mode.} Frequency $\omega_\mathrm{s}$ (a) and decay rate $\gamma$ (b) of the quasi-particles as a function of relative pump power $P/P_\mathrm{cr}$, as extracted from the fit to the dynamic structure factor, Fig. 2. The grey shaded area in the top panel results from an \textit{ab initio} calculation of the expected soft mode frequency, taking into account the experimental uncertainties in the determination of the coherent cavity field and the depth of the optical lattice resulting from the transverse pump field. Close to $P/P_\mathrm{cr}=1$, for $\omega_\mathrm{s}/(2\pi)<400$ Hz, the uncertainty in modeling atom loss leads to a substructure which we omit in the graph. The solid line in the lower panel is a fit with a phenomenological function to the data (see Supplementary Note 3). Vertical and horizontal errorbars indicate the statistical errors (s. d.) reported from the fit, and the error (s.d.) in the determination of the critical point, respectively.}
\label{fig:energygamma}
\end{figure}

\subsection{Characterization of the quasi-particle mode}
The access to the dynamic structure factor $S(\mathbf{k}_\mathrm{cb},\omega)$ allows us to characterize the quasi-particle mode which emerges due to the long-range interactions in the gas. When adiabatically switching on the long-range interactions ($P\neq0$), new quasi-particle modes of polaritonic character form, where intra-cavity photons are admixed to the recoil momentum states. A diagonalization of the Hamiltonian for $P\neq0$ leads to the definition of a quasi-particle mode for the interacting system with annihilation and creation operators $\hat{c}$ and $\hat{c}^\dag$, respectively (Supplementary Note 1).  From our measurements, Fig. 2, we can directly extract the energy $\hbar \omega_\mathrm{s}$ of this quasi-particle mode as a function of $P/P_\mathrm{cr}$. To this end, we fit a resonance curve of a damped harmonic oscillator to both sidebands of $S(\mathbf{k}_\mathrm{cb},\omega)$ (Supplementary Note 2), whose peak positions corresponds to $\omega_\mathrm{s}$, see Fig. 4. We observe the mode softening towards the critical point from both sides of the phase transition. The width $\gamma$ of the sidebands is displayed in the lower panel of Fig. 4 and characterizes the damping of the quasi-particle mode. For our parameters, the main constituent of the quasi-particle mode is the atomic component, while the light field is only weakly admixed. We thus attribute the observed damping mainly to the decay of atomic momentum excitations. The finite decay rate $\kappa$ of the cavity light field gives rise to an additional damping of the quasi-particle mode estimated to be only a few Hertz (Supplementary Note 1). The behavior of the damping rate $\gamma$ has been studied theoretically and originates from a resonant enhancement of the Beliaev damping of the checkerboard density-wave \cite{Kulkarni2013, Konya2014, Konya2014a}. Our characterization of the quasi-particle mode is consistent with earlier measurements \cite{Mottl2012a, Brennecke2013a}, but now also extends into the organized phase because we can distinguish density fluctuations from a density modulation.
\begin{figure}
 \includegraphics[]{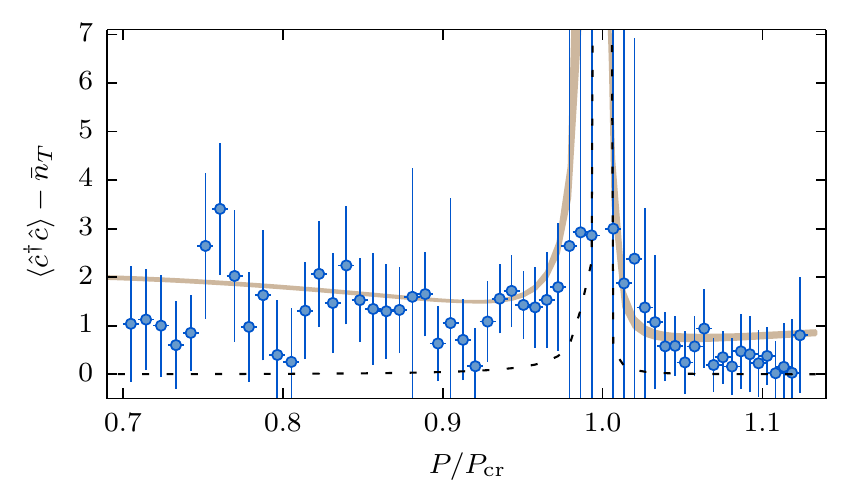}
\caption{\textbf{Number of quasi-particles.} Number of quasi-particles $\bra  \cdag \chat \ket-\bar{n}_T$ as a function of relative pump power $P/P_\mathrm{cr}$, extracted from the sideband asymmetry in the dynamic structure factor. The grey shaded area shows the result from an \textit{ab initio} calculation of the expected quasi-particle mode occupation (Supplementary Notes 1 and 3), taking into account the experimental uncertainties in the determination of the coherent cavity field and the depth of the optical lattice resulting from the transverse pump field.  Shown as black dashed line is the calculated thermal occupation $\bar{n}_T$ of the quasi-particle mode due to the finite temperature of the BEC for a temperature of \unit[38]{nK}. Vertical and horizontal errorbars indicate the statistical error (s. d.) reported from the fit, and the error in the determination of the critical point (s. d.), respectively. The strongly increased vertical errorbars close to $P/P_\mathrm{cr}=1$ arise from the decreasing sideband asymmetry, while their individual errors stay roughly constant. }
 \label{fig:cdagc}
\end{figure}

\subsection{Occupation of the quasi-particle mode}
The occupation $\bra \hat{c}^\dag \hat{c} \ket$ of the quasi-particle mode can be extracted from the observed sideband asymmetry. As can be seen in Fig. 2, the red-shifted sideband dominates over the blue-shifted one. For a system in its ground state, only creation processes are possible, leading to a vanishing blue-shifted sideband. This has been used for thermometry of trapped ions and cavity optomechanical systems \cite{Diedrich1989,Safavi-Naeini2012}. From the observation of the finite blue-shifted sideband in our experiment we infer that the system is in a steady state close to its ground state. The continuous measurement process via cavity decay constantly creates and annihilates quasi-particles at rates $2\kappa \bra \delta\adag \delta\ahat \ket_-$ and $2\kappa \bra \delta\adag \delta\ahat \ket_+$. Here, $\bra \delta\adag \delta\ahat \ket_\pm$ is the integrated spectral weight of the blue ($+$) and red ($-$) sideband, respectively. This measurement backaction effectively gives rise to a heating rate of the system, and thus to a finite occupation of the quasi-particle mode. At the same time,  the finite decay rate $\gamma$ of the quasi-particle mode changes this occupation: On one hand, quasi-particles will be annihilated due to this dissipation channel at rate $2\gamma \bra \cdag \chat \ket$. On the other hand, it couples the quasi-particle mode to a thermal heat bath provided by the atomic cloud. This creates quasiparticles at rate $2\gamma \bar{n}_T$, where $\bar{n}_T$ is the thermal occupation of the quasi-particle mode calculated from the Bose distribution function, and $T=\unit[38(10)]{nK}$ is the temperature of the BEC, measured independently from absorption images. In steady state, the different contributions are balanced according to the rate equation (see Supplementary Note 1 and Supplementary Figure 1),
\begin{equation}
2\kappa\left( \bra \delta\adag \delta\ahat \ket_- - \bra \delta\adag \delta\ahat \ket_+ \right)  = 2\gamma\left( \bra \cdag \chat \ket - \bar{n}_T\right)\,.
\end{equation}

We can determine the occupation $\bra \cdag \chat \ket -\bar{n}_T$ of the quasi-particle mode, using the weights of the sidebands and the dissipation rates $\gamma$ (empirical fit) and $\kappa$ (see Fig. 5). We observe an average occupation of the quasi-particle mode on the level of only a few quanta. In the organized phase an increase of the mode occupation towards the critical point seems visible. The occupation is expected to diverge when approaching the critical point since the energy of the soft mode vanishes and the atomic damping rate goes to zero \cite{Konya2014,Konya2014a}. This situation is very similar to the enhanced thermal occupation of roton-like states predicted for dilute quantum gases with dipolar interactions at finite temperatures \cite{Klawunn2011}. 

\section*{Discussion}
We used vacuum-stimulated scattering of light to directly measure the dynamic structure factor of a quantum gas with cavity-mediated long-range interactions. Access to the dynamic structure factor allowed us to characterize the relevant quasi-particle mode while the system crossed a structural phase transition, to distinguish  density modulation and density fluctuations, and to measure the critical exponents of the density fluctuations. We further extracted the finite occupancy of the quasi-particle mode under the influence of measurement backaction due to cavity decay and an atomic bath at finite temperature. While this measurement was motivated by the very specific setup used to create cavity-mediated long-range interactions, an extension to more general settings seems possible. The approach of applying quantum optical methods based on strong matter-light interaction to the investigation of dilute ultracold gases offers unique possibilities for the non-destructive real-time investigation of quantum matter and its phase transitions \cite{Mekhov2007, DeChiara2011a, Weimer2011, Oztop2012, Roscilde2009}.

\section*{Methods}
After centering an almost pure $^{87}$Rb Bose-Einstein condensate (BEC) trapped in a crossed-beam dipole trap with respect to the TEM$_{00}$ cavity mode, the transverse pump power $P$, at wavelength $\lambda_{\mathrm{p}} = \unit[785.3]{nm}$, is increased over $\unit[100]{ms}$ to a relative coupling strength of $P/P_\mathrm{cr} \approx 0.46$. Subsequently, the power $P$ is linearly increased over $\unit[0.5]{s}$ to $P/P_\mathrm{cr}  \approx 1.38$, while the stream of photons leaking out of the cavity is detected in a balanced heterodyne configuration using a local oscillator (LO) power of $\unit[2.2]{mW}$ and balanced photodiodes (Thorlabs PDB110A) \cite{Baumann2011}. The extracted quadratures at a beat frequency of $\unit[59.55]{MHz}$ are mixed down to $\unit[50] {kHz}$, amplified, low-pass filtered and digitized using high-speed analog-to-digital converters with $\unit[2]{us}$ resolution (National Instruments PCI-6132). The response of the heterodyne system is $\unit[2.2]{V^2~per~cavity~photon}$.

The temperature of the initially prepared BEC was determined from absorption images to be $T = \unit[20(10)]{nK}$. The far-off resonant transverse pump beam heats the BEC during probing to a temperature of $T = \unit[38(10)]{nK}$ at the critical point. Residual atom loss of $26\%$ during probing is included by rescaling the relative coupling axis according to the proportionality $P_\mathrm{cr} \propto N^{-1}$. 

The phase transition point is characterized by a steep increase of the intracavity photon field. We fit the rise of the photon field once it has first exceeded a mean intracavity photon number of $4.5$ with a saturation function $p_0\cdot(1-t_{\mathrm{cr}}/t)^{p_1}$. With the extracted occurence time $t_{\mathrm{cr}}$ of the phase transition, we can convert the time axis into a relative coupling axis $P/P_\mathrm{cr}$. The relative statistical error of $P_\mathrm{cr}$ according to this procedure is given by $5\cdot 10^{-4}$, including intensity fluctuations of the transverse pump.

\subsection{Acknowledgements}
We acknowledge insightful discussions with C. Chin, R. Chitra, S. Diehl, N. Dogra, L. Hruby, and S. Huber. Supported by Synthetic Quantum Many-Body Systems (European Research Council advanced grant), the EU Collaborative Project TherMiQ (Grant Agreement 618074), and the DACH project ``Quantum Crystals of Matter and Light''.

\newpage
\onecolumngrid
\renewcommand{\bibnumfmt}[1]{[S#1]}
\renewcommand{\citenumfont}[1]{S#1}
\section{Supplementary Information}
\subsection{Supplementary Note 1: Theoretical description}
\subsubsection{Relation between the structure factor and the power spectral density in our system}
We start with the definition of the dynamic structure factor as the spatial and temporal Fourier transform of the correlations of the density fluctuations $\delta\hat{\rho}(\mathbf{r},t)$ \cite{Hove1954}
\begin{equation}
S(\mathbf{k},\omega) = \frac{1}{2\pi N/V} \int \mathrm{d}\mathbf{r} \mathrm{d} t \, e^{-i(\mathbf{k r} -\omega t)} \langle \delta\hat{\rho}(\mathbf{r},t)\delta\hat{\rho}(0,0)\rangle \,,
\end{equation}
where $V$ is the volume. With the relation $\rho(\mathbf{r})=\frac{1}{V}\sum_k \rho_\mathbf{k} e^{i\mathbf{kr}}$ this can be rewritten in Fourier space as 
\begin{equation}
S(\mathbf{k},\omega) = \frac{1}{2\pi N} \int \mathrm{d} t \, e^{i\omega t} \langle \delta\hat{\rho}_\mathbf{k}(t)\delta\hat{\rho}_\mathbf{-k}(0)\rangle \,.
\end{equation}

In steady state, the cavity light field operator $\hat{a}$ is given (neglecting small variations in the dispersive shift) by \cite{Brennecke2013a}
\begin{equation}\label{eq:steady_state}
\ahat = \frac{\eta \hat{\Theta}}{\tilde{\Delta}_c-i \kappa}\,,
\end{equation}
where the order parameter $\hat{\Theta} = \Theta_0 + \delta\hat{\Theta}$
\begin{eqnarray}
\hat{\Theta} &=& \Theta_0 + \int \mathrm{d}^3\mathbf{r} \cos{k x}\cos{k z} \delta\hat{\rho}(\mathrm{r},t)\\
 &=& \Theta_0 + \frac{1}{4} \left( \delta\hat{\rho}_{(k,0,k)}  + \delta\hat{\rho}_{(k,0,-k)}  +\delta\hat{\rho}_{(-k,0,k)}  +\delta\hat{\rho}_{(-k,0,-k)}\right)\,.
\end{eqnarray}
is written as a sum of a static density modulation $\Theta_0=\langle \hat{\Theta}\rangle$ and  time-dependent fluctuations. Here, the indices indicate the momentum along the $x$-, $y$-, respectively $z$-axis. We choose the coordinate system such that the cavity mode is oriented along the $x$-direction, and the transverse pump beam propagates along the $z$-direction.

We now expand the cavity light field around its mean field solution $\alpha$, $\hat{a}=\alpha + \delta\hat{a}$ (and correspondingly the order parameter as $\hat{\Theta}=\Theta_0 + \delta\hat{\Theta}$), and use equation (\ref{eq:steady_state}) to find for the correlator
\begin{equation}
\begin{split}
\langle \hat{a}^\dag(t)\hat{a}(0)\rangle&=|\alpha|^2 + \langle \delta\hat{a}^\dag(t) \delta\hat{a}(0)\rangle\\
&= \frac{\eta^2}{\kappa^2 + \tilde{\Delta}_c^2} \left[ \Theta_0^2 + \frac{1}{16} \langle \delta\hat{\rho}_{(k,0,k)} \delta\hat{\rho}_{(-k,0,-k)} + \delta\hat{\rho}_{(k,0,-k)} \delta\hat{\rho}_{(-k,0,k)}\right.\\
&\left.+\delta\hat{\rho}_{(-k,0,k)} \delta\hat{\rho}_{(k,0,-k)} + \delta\hat{\rho}_{(-k,0,-k)} \delta\hat{\rho}_{(k,0,k)} \rangle  \vphantom{\frac12}\right]\\
&=\frac{\eta^2}{\kappa^2 + \tilde{\Delta}_c^2} \left[ \Theta_0^2 + \frac{1}{4} \langle \delta\hat{\rho}_{(k,0,k)} \delta\hat{\rho}_{(-k,0,-k)}\rangle  \right]\,.
\end{split}
\end{equation}
Here, we made use of the fact that $\langle \delta \hat{\rho}_\mathbf{q}(t) \delta \hat{\rho}_\mathbf{q'}(0)\rangle = 0$, unless $\mathbf{q}=-\mathbf{q}'$, and that the system is symmetric under the transformations $x\rightarrow-x$ and $z\rightarrow-z$ due to the involved standing waves. Since the Fourier transform of the temporal correlation function $\langle \hat{a}^\dag(t)\hat{a}(0)\rangle$ is the power spectral density of the intra-cavity light field, we finally obtain a direct relation between the dynamic structure factor and the spectrum of the intra-cavity light field:
\begin{eqnarray}
PSD(\omega) &=& \int \mathrm{d}t\,e^{-i\omega t}\langle \hat{a}^\dag(t)\hat{a}(0)\rangle\\
&=& \frac{\eta^2}{\kappa^2 + \tilde{\Delta}_c^2} \left[  2\pi \Theta_0^2 \delta(\omega) + \frac{2\pi N}{4} S(\mathbf{k}_\mathrm{cb},\omega) \right]\,.
\end{eqnarray}
The power spectral density $PSD(\omega)$ thus quantifies both the checkerboard density modulation (first term, zero frequency bin) and the dynamic structure factor at the wavevector $\mathbf{k}_\mathrm{cb} = (\pm k,0, \pm k)$ (second term).

Finally we can extract the dynamic structure factor as
\begin{equation}
  S(\mathbf{k}_\mathrm{cb},\omega) =  \frac{\kappa^2+\tilde{\Delta}_c^2}{\eta^2} \frac{4}{N}  \left( \frac{1}{2\pi} PSD(\omega) - |\alpha|^2 \delta({\omega})\right)\,.
\end{equation}

For the experimentally relevant case of two nearly degenerate, circularly polarized cavity modes, we find accordingly
\begin{equation}
S(\mathbf{k}_\mathrm{cb},\omega) =  \frac{\kappa^2+\tilde{\Delta}^\textrm{(eff)\,2}_c}{\eta_1^2+\eta_2^2} \frac{4}{N}  \left( \frac{1}{2\pi} PSD(\omega) - |\alpha|^2 \delta({\omega})\right)\,,
\end{equation}
where the definitions of $\eta_i$ and $\tilde{\Delta}_{c,\textrm{eff}}$ are given below, and $\alpha$ and $PSD(\omega)$ refer to the according superposition of the bare cavity modes coupling to the atoms.

\subsubsection{Derivation of the Hamiltonian describing the fluctuations of the system}
In this section we derive an effective Hamiltonian describing the fluctuations of the system starting from the many-body Hamiltonian,
\begin{equation} \label{eq:man-body-H}
\hat{H}_\mathrm{mb} = \hat{H}_\textrm{c} +\hat{H}_\textrm{a} +\hat{H}_\textrm{a-c} +\hat{H}_\textrm{SB}\,, 
\end{equation}
with 
\begin{equation}
\begin{split}
&\hat{H}_\textrm{c} = -\hbar \Delta_\textrm{c} \sum_{i=1}^2 \hat{a}^\dag_i \hat{a}_i \\
&\hat{H}_\textrm{a} = \int \mathrm{d}^3 r \hat{\Psi}^\dag (\mathbf{r}) \left[ \frac{\mathbf{p}^2}{2m} + V_\mathrm{p} \cos^2(kz) + \frac{g}{2}\hat{\Psi}^\dag(\mathbf{r})\hat{\Psi}(\mathbf{r}) \right] \hat{\Psi}(\mathbf{r})\\
&\hat{H}_\textrm{a-c} = \sum_{i=1}^2 \int \mathrm{d}^3 r \hat{\Psi}^\dag (\mathbf{r}) \left[ \hbar \eta_i \cos(kx) \cos(kz) (\hat{a}_i + \hat{a}_i^\dag) + \hbar U^i_0 \cos^2(kx) \hat{a}_i^\dag \hat{a}_i \right] \hat{\Psi}(\mathbf{r})\\
&\hat{H}_\textrm{SB} = \hbar \zeta \sum_{i=1}^2 \eta_i (\hat{a}_i + \hat{a}_i^\dag)\,.
\end{split}
\end{equation}
Here we take into account that the cavity is supporting two degenerate (neglecting a small birefringence on the order of  the cavity line width), circularly polarized cavity modes with annihilation operators $(\hat{a}_1, \hat{a}_2)$, which are coupled to the atoms with two-photon Rabi frequencies $(\eta_1, \eta_2)$ \cite{Mottl2012a}. The maximum dispersive shift of the two cavity modes due to the dispersive coupling of a single atom is described by $(U_0^1, U_0^2)$. A symmetry breaking term $\hat{H}_\textrm{SB}$ is introduced, which is proportional to a real-valued effective cavity drive amplitude $\zeta$ (see also definitions in \cite{Brennecke2013a}).

Following references \cite{Nagy2008, Mottl2012a}, we expand the atomic and cavity field operators ($\hat{\Psi}, \hat{a_i}$) around their mean-field values ($\psi_0, \alpha_0^i$), 
\begin{align}\label{eq:steady-state-expansion}
&\hat{\Psi}=(\sqrt{N}\psi_0+\delta\hat{\Psi})e^{-it\mu_0/\hbar}\\
&\hat{a}_i=\alpha_0^i+\delta\hat{a}_i\,.
\end{align}

Expanding Hamiltonian $\hat{H}_\mathrm{mb}$, Eq. (\ref{eq:man-body-H}), in the limit $|\Delta_c|\gg\kappa$ up to second order in the fluctuation operators ($\delta\hat{\Psi}, \delta\hat{a}_i$), we find the quadratic Hamiltonian
\begin{equation}
\hat{H}^{(2)} = \hat{H}_0 + \hbar \sum_{i=1}^2 \left(\eta_i \delta \hat{\Theta} + U_0^i \alpha_0^i \delta\hat{B}\right)(\delta\hat{a}_i + \delta\hat{a}_i^\dag) + \hbar\tilde{\Delta}_c^i \delta\hat{a}_i^\dag\delta\hat{a}_i\,,
\end{equation}
with 
\begin{equation}
\begin{split}
\hat{H}_0 = & \int d^3r \,\,\delta\hat{\Psi}^\dag \left[ \frac{-\hbar^2}{2m}(\partial^2_x+\partial^2_z) +V_\mathrm{p}(z)+\hbar\sum_{i=1}^2\eta_i(x,z) (\alpha_0^i+\alpha_0^{i*}) \right.\\
&\left.+ \hbar \sum_{i=1}^2 U_0^i(x) |\alpha_0^i|^2 \right] \delta\hat{\Psi}+\frac{1}{2}\,g_\mathrm{2D}\,\psi_0^2\left(\delta\hat{\Psi}^2+(\delta\hat{\Psi}^\dag)^2\right)
+ 2g_\mathrm{2D}|\psi_0|^2\delta\hat{\Psi}^\dag\delta\hat{\Psi}\,,
\end{split}
\end{equation}
where $\tilde{\Delta}_c^i = \Delta_c - \mathcal{B}_0 U_0^i$. We introduced here the spatially dependent  classical lattice potential $V_p(z)=V_p\cos^2(kz)$ of the transverse pump, the spatially dependent two-photon Rabi frequencies $\eta_i(x,z) = \frac{\Omega_p g_0^{(i)}}{\Delta_a}\cos(k x)\cos(k z)$ with maximum pump Rabi frequency $\Omega_p$, detuning $\Delta_a=\omega_p-\omega_a$ between atomic resonance frequency and pump light frequency, and single-atom coupling strengths $g_0^{(i)}$. Further, the light shift per photon is $U_0^i(x)=\frac{g_0^{(i)\,2}}{\Delta_a}\cos^2(k x)$ . Along the third direction, $y$, we assume a homogeneous system and use the according contact interaction strength $g_{2D}$ \cite{Kramer2003}. We introduced the definition $\delta\hat{\Theta}=\sqrt{N}\int\mathrm{d}^3 r (\delta\hat{\Psi}^\dag + \delta\hat{\Psi}) \cos(kx)\cos(kz)\psi_0$ of the fluctuations of the order parameter $\hat{\Theta}=\int\mathrm{d}^3 r\hat{\Psi}^\dag \hat{\Psi} \cos(kx)\cos(kz)$ around the mean field value $\Theta_0=\int\mathrm{d}^3 r |\psi_0|^2 \cos(kx)\cos(kz)$. In a similar way, we define the fluctuation operator  $\delta\hat{\mathcal{B}}=\sqrt{N}\int\mathrm{d}^3 r (\delta\hat{\Psi}^\dag + \delta\hat{\Psi}) \cos(kx)^2\psi_0$ of the bunching operator $\hat{\mathcal{B}}=\int\mathrm{d}^3 r\hat{\Psi}^\dag \hat{\Psi} \cos(kx)^2$ around its mean-field value $\mathcal{B}_0=\int\mathrm{d}^3 r |\psi_0|^2 \cos(kx)^2$.

The mean-field values are determined by the corresponding Gross-Pitaevskii equation derived from $\hat{H}^{(2)}$:
\begin{equation}
\begin{split}
\mu_0 \psi_0 =& \left( -\frac{\hbar^2}{2m} (\partial_x^2 + \partial_z^2) +V_p(z) + \hbar\sum_{i=1}^2 U_0^i(x)
|\alpha_0^i|^2 \right.\\ 
&\left.+\hbar \sum_{i=1}^2 \eta_i(x,z) (\alpha_0^i+\alpha_0^{i*} + g_{2D}|\psi_0|^2)\right) \psi_0(x,z)\,,
\end{split}
\end{equation}
with $\alpha_0^i = \frac{\eta_i \Theta_0}{\tilde{\Delta_c^i}+ i\kappa}$, assuming adiabatic following of the cavity field.

To find the collective excitations of the system around the mean-field solution, we expand $\delta\hat{\Psi}$ in Bogoliubov modes $\hat{h}_j$ (which diagonalize $H_0$ with eigenenergies $E_j$) with amplitudes $u_j(\mathbf{r})$ and $v_j(\mathbf{r})$,
\begin{equation}
\delta\hat{\Psi}(\mathbf{r}) = \sum_{j} \left(
u_{j}(\mathbf{r})\hat{h}_{j} + v_{j}^*(\mathbf{r})\hat{h}_{j}^\dag
\right)\,,
\end{equation}
resulting in 
\begin{equation}
 \begin{split}
  \hat{H}^{(2)} =& \sum_j \left[E_j \hat{h}_j^\dag \hat{h}_j + \hbar\sum_{i=1}^2\eta_i\left(\hat{h}_j \sqrt{N \chi_j^{i*}}+ \textrm{h.c.} \right) \left(\delta\hat{a}_i^\dag+\delta\hat{a}_i \right)\right] \\
 &+\hbar\sum_{i=1}^2 -\tilde{\Delta}_c^i \delta\hat{a}_i^\dag\delta\hat{a}_i\,.
\end{split}
\end{equation}
Here, we introduced interaction matrix elements $\chi_j^i=\langle \psi_0 | \cos(kx)\cos(kz) +\frac{\Theta_0 U_0^i}{\tilde{\Delta}_c^i}\cos^2(kx)|u_j+v_j\rangle$, describing the overlap between the ground state wave function and the Bogoliubov excitations as presented in reference \cite{Mottl2012a}. From numerical calculations we know that only a single Bogoliubov mode $\hat{h}_0 = \hat{h}$ with energy $E_0 = E$ is dominantly contributing via the matrix elements $\chi^i_0 = \chi^i$, and all other matrix elements are suppressed by more than two orders of magnitude \cite{Mottl2012a}. This leads to
\begin{equation}
  \hat{H}^{(2)} = E\, \hat{h}^\dag \hat{h} + \hbar \sum_{i=1}^2 -\tilde{\Delta}_c^i \, \delta\hat{a}_i^\dag\delta\hat{a}_i +  \hbar \sum_{i=1}^2\eta_i\sqrt{N}\left(\hat{h} \sqrt{\chi^{i*}} + \hat{h}^\dag \sqrt{\chi^i}\right) \left(\delta\hat{a}_i^\dag+\delta\hat{a}_i \right)\,.
\end{equation}

If we now use the definitions $\omega_0=E/\hbar$ for the bare energy of the uncoupled system, $\lambda_i=\sqrt{N}\eta_i \sqrt{|\chi^i|}$ for the coupling strength and $\delta\hat{b}=\hat{h}e^{i\phi_i}$ for the collective atomic fluctuations (using $\sqrt{\chi^i} = \sqrt{|\chi^i|}e^{i\phi_i} \approx \sqrt{|\chi^i|}$, where $\phi_i \approx 0$, i.e. $\phi_1\approx \phi_2$), we finally arrive at the fluctuation Hamiltonian
\begin{equation}
  \hat{H}=  \hbar \omega_0\delta\hat{b}^\dag\delta\hat{b} + \hbar \sum_{i=1}^2 -\tilde{\Delta}_c^i \delta\hat{a}_i^\dag\delta\hat{a}_i + \hbar \sum_{i=1}^2\lambda_i \left(\delta\hat{a}_i+\delta\hat{a}_i^\dag\right) \left(\delta\hat{b} + \delta\hat{b}^\dag\right)\,.
\end{equation}
In order to further simplify this expression, we apply a transformation which reduces the Hamiltonian describing two cavity modes to a Hamiltonian involving coupling to only one effective cavity mode. We introduce two new cavity modes with fluctuation operators $(\hat{o}_1, \hat{o}_2)$, which are a linear superposition of the original modes $(\delta\hat{a}_1, \delta\hat{a}_2)$
\begin{eqnarray}
  \delta\hat{a}_1 &=& A \hat{o}_1 + B \hat{o}_2\\
  \delta\hat{a}_2 &=& C \hat{o}_1 + D \hat{o}_2\,.
\end{eqnarray}

Applying this transformation and requesting that the new operators obey bosonic commutation relations ($[\hat{o}_1,\hat{o}_1^\dag] = 1, [\hat{o}_2,\hat{o}_2^\dag] = 1$), that mixed terms of the new operators vanish, and that mode $\hat{o}_2$ decouples from the atoms, we find the coefficients
\begin{equation}
  A = \frac{\lambda_1 \tilde{\Delta}_c^{(2)}}{\lambda_2 \tilde{\Delta}_c^{(1)}} C\,, \quad 
B = \frac{\sqrt{\left( \frac{\lambda_1 \tilde{\Delta}_c^{(2)}}{\lambda_2 \tilde{\Delta}_c^{(1)}}\right)^2 +1}}{1+\frac{\lambda_1^2 \tilde{\Delta}_c^{(2)}}{\lambda_2^2 \tilde{\Delta}_c^{(1)}}}\,, \quad
C = -\frac{\lambda_1}{\lambda_2}B\,, \quad 
D = \frac{\sqrt{\left( \frac{\lambda_1}{\lambda_2}\right)^2 +1}}{1+\frac{\lambda_1^2 \tilde{\Delta}_c^{(2)}}{\lambda_2^2 \tilde{\Delta}_c^{(1)}}}\,.
\end{equation}

Inserting these coefficients into $\hat{H}$ yields
\begin{equation}
\begin{split} 
\hat{H} =& -\hbar \frac{(\lambda_1^2 + \lambda_2^2) \tilde{\Delta}_c^{(1)}\tilde{\Delta}_c^{(2)} }{\lambda_1^2 \tilde{\Delta}_c^{(2)} + \lambda_2^2 \tilde{\Delta}_c^{(1)}} \hat{o}_1^\dag \hat{o}_1 -\hbar \frac{\lambda_1^2 \tilde{\Delta}_c^{(2)\,2} + \lambda_2^2 \tilde{\Delta}_c^{(1)\,2} }{\lambda_1^2 \tilde{\Delta}_c^{(2)} + \lambda_2^2 \tilde{\Delta}_c^{(1)}} \hat{o}_2^\dag \hat{o}_2 \\
&+ \hbar\sqrt{\lambda_1^2 +  \lambda_2^2} (\hat{o}_1+\hat{o}_1^\dag)(\delta \hat{b} + \delta \hat{b}^\dag) + \hbar \omega_0 \delta\hat{b}^\dag \delta\hat{b}\,.
\end{split}
\end{equation}
The mode $\hat{o}_2$ is now decoupled from atomic motion and we can write the effective fluctuation Hamiltonian for the system
\begin{eqnarray}\label{eq:Hfluc}
\hat{H}/\hbar=\omega_0 \delta\hat{b}^\dag \delta\hat{b} - \tilde{\Delta}_{c,\textrm{eff}} \delta\hat{a}^\dag \delta\hat{a} +\lambda_\textrm{eff}(\delta\hat{a}^\dag + \delta\hat{a})(\delta\hat{b}^\dag + \delta\hat{b})\,,
\end{eqnarray}
where we used
\begin{eqnarray}
  \delta \hat{a} &=& \hat{o}_1 \\
  \tilde{\Delta}_{c,\textrm{eff}} &=& \frac{(\lambda_1^2 + \lambda_2^2) \tilde{\Delta}_c^{(1)}\tilde{\Delta}_c^{(2)} }{\lambda_1^2 \tilde{\Delta}_c^{(2)} + \lambda_2^2 \tilde{\Delta}_c^{(1)}} \\
  \lambda_\textrm{eff} &=& \sqrt{\lambda_1^2 + \lambda_2^2}\,.
\end{eqnarray}

\subsubsection{Langevin description of the coupled system}
The goal of this section is to derive coupled quantum Langevin equations effectively describing our system. The hierarchy of the relevant energy scales determines their derivation. The fastest timescale in the system is given by the cavity field decay at rate $\kappa \approx 2\pi \times 1.25$ MHz, followed by the variable coupling rate $\lambda_\textrm{eff}$, which reaches $\approx 2\pi \times  100$ kHz at the critical point, and finally the damping rate $\gamma$ of the atomic excitation, which is on the order of a few hundred Hertz \cite{Brennecke2013a}. Following this hierarchy, we first couple the damped optical cavity mode to the Bogoliubov mode $\delta \hat{b}$. This results in new polariton modes of the system which describe the quasi-particles of the long-range interacting system. Only thereafter we introduce the damping of the atomic polariton mode at rate $\gamma$. This approach is justified as the admixture of the cavity mode to the polariton is very small ($\omega_0/\tilde{\Delta}_{c,\textrm{eff}} \ll1$).

We start with Hamiltonian Eq. (\ref{eq:Hfluc}),

\begin{equation}
\hat{H}/\hbar=\omega_0 \delta\hat{b}^\dag \delta\hat{b} + \tilde{\Delta}_{c,\textrm{eff}} \delta\hat{a}^\dag \delta\hat{a} +\lambda_\textrm{eff}(\delta\hat{a}^\dag + \delta\hat{a})(\delta\hat{b}^\dag + \delta\hat{b})
\end{equation}

The corresponding quantum Langevin equation with cavity field damping rate $\kappa$ read

\begin{eqnarray}\label{eq:Langevin1}
\delta\dot{\hat{a}} &=& (-i \tilde{\Delta}_{c,\textrm{eff}} -\kappa)\delta\hat{a} - i\lambda_\textrm{eff}(\delta\hat{b}+\delta\hat{b}) + \sqrt{2\kappa} \ain\\ 
\delta\dot{\hat{b}} &=& -i\omega_0 \delta\hat{b} - i\lambda_\textrm{eff}(\delta\hat{a}+\delta\hat{a}^\dag)
\end{eqnarray}

The bosonic operator $\ain$  in Eq. \eqref{eq:Langevin1}, describes vacuum input fluctuations of the surrounding electromagnetic field modes which are characterized by the correlation functions $\langle \ain(t)\ain^\dag(t')\rangle = \delta(t-t')$ and $\langle \ain^\dag(t)\ain(t')\rangle = 0$.

In order to find the new eigenmodes (polaritons) of the coupled system we rewrite the Langevin equations in matrix form,
\begin{equation}\label{eq:Langevin_matrix_bare}
\frac{d}{dt} 
\begin{pmatrix} \delta\ahat \\ \delta\adag \\ \delta\bhat \\ \delta\bdag \end{pmatrix} 
= M_0 \begin{pmatrix} \delta\ahat \\ \delta\adag \\ \delta\bhat \\ \delta\bdag \end{pmatrix} 
+\sqrt{2\kappa}
\begin{pmatrix} \ain \\ \ain^\dag \\ 0 \\ 0 \end{pmatrix} 
\end{equation}
with the matrix
\begin{equation}
M_0=
\begin{pmatrix}
-i\tilde{\Delta}_{c,\textrm{eff}}-\kappa & 0 & -i\lambda_\textrm{eff} & -i\lambda_\textrm{eff} \\ 
0 & i\tilde{\Delta}_{c,\textrm{eff}}-\kappa & i\lambda_\textrm{eff} & i\lambda_\textrm{eff} \\ 
-i\lambda_\textrm{eff} & -i\lambda_\textrm{eff} & -i\omega_0 & 0 \\ 
i\lambda_\textrm{eff} & i\lambda_\textrm{eff} & 0 & -i\omega_0 
\end{pmatrix} 
\end{equation}

Diagonalization of $M_0$ via the transformation
\begin{equation}
S^{-1}M_0 S = D = \begin{pmatrix}
-i\tilde{\Delta}_{c,\textrm{eff}}-\kappa & 0 & 0 & 0 \\ 
0 & i\tilde{\Delta}_{c,\textrm{eff}}-\kappa & 0 & 0 \\ 
0 & 0 & -i\omega_s & 0 \\ 
0 & 0 & 0 & i\omega_s 
\end{pmatrix} 
\end{equation}
with, neglecting terms of order $\frac{\omega_0}{\omega}$,
\begin{equation}
S = \begin{pmatrix}
1 & 0 & \frac{-\lambda_\textrm{eff}}{\tilde{\Delta}_{c,\textrm{eff}}-i\kappa}\sqrt{\frac{\omega_0}{\omega_s}} & \frac{-\lambda_\textrm{eff}}{\tilde{\Delta}_{c,\textrm{eff}}-i\kappa}\sqrt{\frac{\omega_0}{\omega_s}} \\ 
0 & 1 & \frac{-\lambda_\textrm{eff}}{\tilde{\Delta}_{c,\textrm{eff}}+i\kappa}\sqrt{\frac{\omega_0}{\omega_s}} & \frac{-\lambda_\textrm{eff}}{\tilde{\Delta}_{c,\textrm{eff}}+i\kappa}\sqrt{\frac{\omega_0}{\omega_s}} \\ 
\frac{-i \lambda_\textrm{eff}}{i\tilde{\Delta}_{c,\textrm{eff}}+\kappa} & \frac{-i \lambda_\textrm{eff}}{-i\tilde{\Delta}_{c,\textrm{eff}}+\kappa} & \frac{1+\omega_s/\omega_0}{2}\sqrt{\frac{\omega_0}{\omega_s}} & \frac{1-\omega_s/\omega_0}{2}\sqrt{\frac{\omega_0}{\omega_s}} \\ 
\frac{i \lambda_\textrm{eff}}{i\tilde{\Delta}_{c,\textrm{eff}}+\kappa} & \frac{i \lambda_\textrm{eff}}{-i\tilde{\Delta}_{c,\textrm{eff}}+\kappa} & \frac{1-\omega_s/\omega_0}{2}\sqrt{\frac{\omega_0}{\omega_s}} & \frac{1+\omega_s/\omega_0}{2}\sqrt{\frac{\omega_0}{\omega_s}}
\end{pmatrix} 
\end{equation}
and 
\begin{equation}
S^{-1} = \begin{pmatrix}
1 & 0 & \frac{i\lambda_\textrm{eff}}{i\tilde{\Delta}_{c,\textrm{eff}}+\kappa} & \frac{i\lambda_\textrm{eff}}{i\tilde{\Delta}_{c,\textrm{eff}}+\kappa}\\ 
0 & 1 & \frac{-i\lambda_\textrm{eff}}{-i\tilde{\Delta}_{c,\textrm{eff}}+\kappa} & \frac{-i\lambda_\textrm{eff}}{-i\tilde{\Delta}_{c,\textrm{eff}}+\kappa} \\ 
\frac{- \lambda_\textrm{eff}}{\tilde{\Delta}_{c,\textrm{eff}}-i\kappa} \sqrt{\frac{\omega_0}{\omega_s}} & \frac{\lambda_\textrm{eff}}{\tilde{\Delta}_{c,\textrm{eff}}+i\kappa}\sqrt{\frac{\omega_0}{\omega_s}} & \frac{1+\omega_0/\omega_s}{2}\sqrt{\frac{\omega_s}{\omega_0}} & \frac{1-\omega_0/\omega_s}{2}\sqrt{\frac{\omega_s}{\omega_0}}\\ 
\frac{\lambda_\textrm{eff}}{\tilde{\Delta}_{c,\textrm{eff}}-i\kappa} \sqrt{\frac{\omega_0}{\omega_s}} & \frac{-\lambda_\textrm{eff}}{\tilde{\Delta}_{c,\textrm{eff}}+i\kappa}\sqrt{\frac{\omega_0}{\omega_s}} & \frac{1-\omega_0/\omega_s}{2}\sqrt{\frac{\omega_s}{\omega_0}} & \frac{1+\omega_0/\omega_s}{2}\sqrt{\frac{\omega_s}{\omega_0}}
\end{pmatrix} 
\end{equation}
allows us to define the polariton mode operators $\dhat$ and $\chat$. Here, 

\begin{eqnarray}\label{eq:omega_s}
  \omega_s = \omega_0\sqrt{1-(\lambda_\textrm{eff}/\lambda_\textrm{eff, cr})^2}
\end{eqnarray}
is the eigenfrequency of the quasi-particle mode we are interested in \cite{Nagy2011}. Note that $S^{-1}$ is normalized such that the bosonic commutation relations for $\dddag$ and $\cdag$ defined as
\begin{equation}
\begin{pmatrix} \dhat \\ \dddag \\ \chat \\ \cdag \end{pmatrix} 
= S^{-1}
\begin{pmatrix} \delta\ahat \\ \delta\adag \\ \delta\bhat \\ \delta\bdag \end{pmatrix} 
\end{equation}

are valid. This way we achieve the following expressions for the operators $\dhat$ ($\chat$), annihilating an excitation in the optical (atomic) polariton mode with eigenfrequency $\tilde{\Delta}_{c,\textrm{eff}}$ ($\omega_s$):

\begin{equation}
\begin{split}
\dhat & =  \delta\ahat + \frac{i\lambda_\textrm{eff}}{i\tilde{\Delta}_{c,\textrm{eff}} + \kappa}(\delta\bhat+\delta\bdag)\\
\chat  & =  \sqrt{\frac{\omega_0}{\omega_s}}\lambda_\textrm{eff} \left(\frac{-\delta\ahat}{\tilde{\Delta}_{c,\textrm{eff}}-i\kappa} + \frac{\delta\adag}{\tilde{\Delta}_{c,\textrm{eff}}+i\kappa}\right) \\ 
& + \frac{1}{2}\sqrt{\frac{\omega_s}{\omega_0}} \left((1+\sqrt{\frac{\omega_0}{\omega_s}})\delta\bhat + (1-\sqrt{\frac{\omega_0}{\omega_s}})\delta\bdag \right).
\end{split}
\end{equation}

Again, this is correct up to terms of order $\omega_0/\tilde{\Delta}_{c,\textrm{eff}}$. For $\lambda_\textrm{eff} \rightarrow 0$ these equations  yield $\dhat \rightarrow \delta\ahat$ and $\chat \rightarrow \delta\bhat$, such that the polariton modes evolve into the bare modes for vanishing coupling.

\subsubsection{Langevin equations for the polariton modes}
We now can write down the Langevin equations for the polariton modes. Starting from equation (\ref{eq:Langevin_matrix_bare}) we use

\begin{equation}
\frac{d}{dt} \begin{pmatrix} \dhat \\ \dddag \\ \chat \\ \cdag \end{pmatrix} 
= \frac{d}{dt} S^{-1}
\begin{pmatrix} \delta\ahat \\ \delta\adag \\ \delta\bhat \\ \delta\bdag \end{pmatrix} 
= S^{-1}M_0S S^{-1} \begin{pmatrix} \delta\ahat \\ \delta\adag \\ \delta\bhat \\ \delta\bdag \end{pmatrix} 
+\sqrt{2\kappa}S^{-1}
\begin{pmatrix} \ain \\ \ain^\dag \\ 0 \\ 0 \end{pmatrix} 
\end{equation}

to find

\begin{eqnarray}\label{eq:Langevin}
\dot{\hat{d}} &=& (-i \tilde{\Delta}_{c,\textrm{eff}} -\kappa)\hat{d}+ \sqrt{2\kappa} \ain\\ 
\dot{\hat{c}} &=& -i\omega_s \hat{c} + \sqrt{2\kappa} \sqrt{\frac{\omega_0}{\omega_s}}\lambda_\textrm{eff}\left( \frac{-\ain}{\tilde{\Delta}_{c,\textrm{eff}}-i\kappa} + \frac{\ain^\dag}{\tilde{\Delta}_{c,\textrm{eff}}+i\kappa}\right) \label{eq:c_mode_undamped}
\end{eqnarray}

The input noise terms in the equation for $\chat$ correspond to the quantum backaction: although the cavity bath is at $T=0$, the $c$-mode is driven by quantum noise from the open channel.

\subsubsection{Damping of the $c$-mode}
We now formally introduce a damping of the atomic polariton mode which effectively models its collisional interaction with the surrounding Bogoliubov modes of the atomic cloud. In principle, this mode is also damped by the cavity decay channel. However, since $\omega_0/\tilde{\Delta}_{c,\textrm{eff}} \ll 1$, this damping rate is on the order of a few Hertz and will be neglected here. We define input noise operators with 
correlation functions $\langle \cin(t)\cin^\dag(t')\rangle = \delta(t-t')(1+\bar{n}_T)$ and $\langle \cin^\dag(t)\cin(t')\rangle = \delta(t-t')\bar{n}_T$. The thermal occupation number $\bar{n}_T$ of the atomic polariton mode is modeled by the Bose distribution function, evaluated at the soft mode frequency $\omega_s$ of the coupled system. Accordingly, the Langevin equation (\ref{eq:c_mode_undamped}) becomes

\begin{equation}\label{eq:equationofmotion_c}
\dot{\chat}=(-i\omega_s-\gamma)\chat + \sqrt{2\gamma}\cin + \sqrt{2\kappa} \sqrt{\frac{\omega_0}{\omega_s}}\lambda_\textrm{eff}\left( \frac{-\ain}{\tilde{\Delta}_{c,\textrm{eff}}-i\kappa} + \frac{\ain^\dag}{\tilde{\Delta}_{c,\textrm{eff}}+i\kappa}\right)
\end{equation}

\subsubsection{Occupation number of the $c$-mode}
With this result, we can now calculate the expectation value $\bra \cdag(t) \chat(t) \ket$ for the number of quasi-particles. From the solution of Eq. (\ref{eq:equationofmotion_c}),
\begin{equation}
\chat(t) = \int_0^t e^{(-i\omega_s-\gamma)(t-t')} \left(\sqrt{2\gamma}\cin +\sqrt{2\kappa} \sqrt{\frac{\omega_0}{\omega_s}}\lambda_\textrm{eff}\left( \frac{-\ain}{\tilde{\Delta}_{c,\textrm{eff}}-i\kappa} + \frac{\ain^\dag}{\tilde{\Delta}_{c,\textrm{eff}}+i\kappa}\right)\right)\,,
\end{equation}
we find 
\begin{equation} \label{eq:cdagc}
\bra \cdag(t) \chat(t) \ket = \bar{n}_T + \frac{2\kappa}{2\gamma}\frac{\omega_0}{\omega_s}\frac{\lambda_\textrm{eff}^2}{\tilde{\Delta}_{c,\textrm{eff}}^2+\kappa^2}\,,
\end{equation}
which diverges towards the critical point with an exponent 0.5 as a function of $1-\lambda_\textrm{eff}/\lambda_\textrm{cr}$.

\subsubsection{Spectrum of the light field}
The main observable in our experiment is the photon spectrum of the light field leaking out of the cavity. In order to find a relation between this spectrum and the $\chat$- and $\dhat$-mode, we apply a backtransformation using the matrix $S$, and move to frequency space (for the definition of the Fourier transformation see \cite{Brennecke2013a}):

\begin{eqnarray}
\delta\ahat(\omega) & = & \dhat(\omega) - \lambda_\textrm{eff} \frac{\tilde{\Delta}_{c,\textrm{eff}}+ i \kappa}{\tilde{\Delta}_{c,\textrm{eff}}^2+\kappa^2} \sqrt{\frac{\omega_0}{\omega_s}} \left(\chat(\omega)+\cdag(-\omega)\right) \\
\delta\adag(\omega) & = & \dddag(\omega) - \lambda_\textrm{eff} \frac{\tilde{\Delta}_{c,\textrm{eff}} + i \kappa}{\tilde{\Delta}_{c,\textrm{eff}}^2+\kappa^2} \sqrt{\frac{\omega_0}{\omega_s}} \left(\cdag(\omega)+\chat(-\omega)\right) 
\end{eqnarray}

Using the Langevin equation in Fourier space \cite{Dimer2007}, the expectation value of the cavity output photon spectrum can be written as
\small
\begin{equation} \label{eq:spectrum}
\begin{split}
\bra \delta\adag(\omega) \delta\ahat(\omega') \ket & = \bra \dddag(\omega) \dhat(\omega') \ket + \frac{\lambda_\textrm{eff}^2}{\tilde{\Delta}_{c,\textrm{eff}}^2+\kappa^2}\frac{\omega_0}{\omega_s}\left( \bra \cdag(\omega) \chat(\omega') \ket + \bra \chat(-\omega) \cdag(-\omega') \ket\right)\\
&= \frac{\lambda_\textrm{eff}^2}{\tilde{\Delta}_{c,\textrm{eff}}^2+\kappa^2}\frac{\omega_0}{\omega_s} \left( \frac{2\gamma \delta(\omega-\omega') \bar{n}_T + \frac{2\kappa \lambda_\textrm{eff}^2}{\tilde{\Delta}_{c,\textrm{eff}}^2+\kappa^2}\frac{\omega_0}{\omega_s} \delta(\omega-\omega')}{|i\omega_s+\gamma-i\omega|^2}\right.\\
&+\left.\frac{2\gamma \delta(\omega-\omega') (1+\bar{n}_T) + \frac{2\kappa \lambda_\textrm{eff}^2}{\tilde{\Delta}_{c,\textrm{eff}}^2+\kappa^2}\frac{\omega_0}{\omega_s} \delta(\omega-\omega')}{|i\omega_s+\gamma+i\omega|^2}\right)
\end{split}
\end{equation}
\normalsize

We set $\bra \dddag(\omega) \dhat(\omega') \ket=0$, as the photonic polariton mode occupation is vanishingly small. In the last line of equation (\ref{eq:spectrum}), the first term in the sum corresponds to the blue-shifted sideband, while the second term corresponds to the red-shifted sideband. The asymmetry in the amplitude of the sidebands is thus given by the different factors $\bar{n}_T$ and $(1+\bar{n}_T)$. The expectation value $\bra \delta\adag(\omega) \delta\ahat(\omega') \ket$ corresponds to the experimentally observed power spectral density $PSD(\omega)$, excluding the coherent part at $\omega=0$.

\renewcommand{\figurename}[2]{Supplementary Figure 1} 
\begin{figure}[ht]
 \includegraphics[width = 10cm]{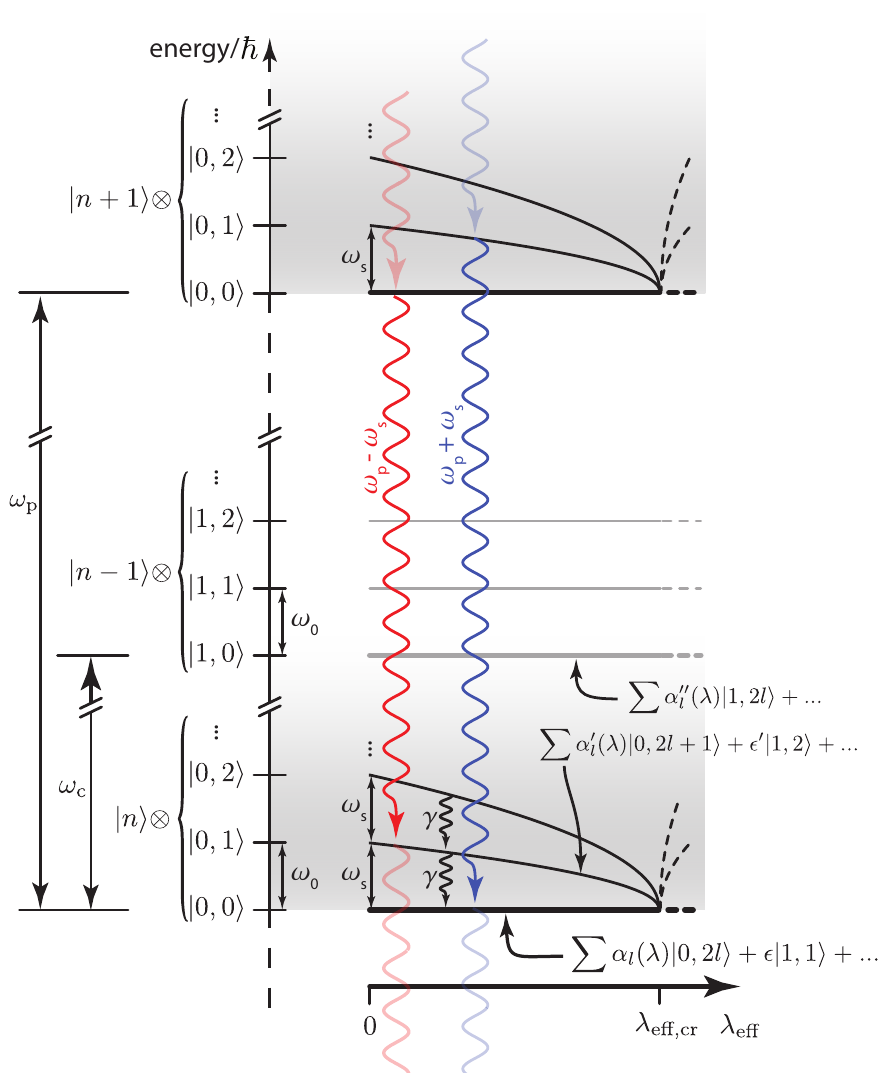}
\caption{\setstretch{1.0} \textbf{Energy diagram of the driven BEC-cavity system.} The bare states are denoted by
$|n_p\ket\otimes|n_c,n_a\ket$, where $n_p$ is the photon number of the coherent
transverse pump field (shown are the manifolds of states for two different $n_p$), $n_c$
is the intra-cavity photon number and $n_a$ is the number of atoms with
$(p_x,p_z)=(\pm\hbar k_x, \pm\hbar k_z)$. The coupling which conserves parity of
$n_a+n_c$ softens the atomic excitation spectrum for increasing coupling strength $\lambda_\textrm{eff}$.
Admixtures of states to the coupled states are indicated. At the critical coupling strength
$\lambda_\textrm{eff, cr}$, the excitation gap $\hbar \omega_s$ vanishes, triggering the quantum
phase transition to a self-organized atomic state. The finite cavity decay makes this phase
transition of non-equilibrium nature as the corresponding decay processes (wiggling lines) give
rise to a depletion of the ground state of the Hamiltonian system. The exiting cavity photons
at frequency $\omega_p\pm\omega_s$ carry real-time information about the dynamics of
the quantum many-body system, corresponding to the creation (red line) and annihilation (blue line) of quasi-particles. The collective momentum excitations can decay at rate $\gamma$, which results in an effective decrease of the annihilation processes via photon decay. This leads to the observed sideband asymmetry.}
 \label{fig:levelscheme}
\end{figure}

\subsubsection{Sideband asymmetry}
We can now explicitly calculate the sideband asymmetry in the cavity output photon spectrum, defined as the difference between the spectral weight of the red-shifted and the blue-shifted sideband. With the definition of the mean intracavity photon number in steady state \cite{Dimer2007}
\begin{eqnarray}
  \langle \hat{a}^\dag \hat{a} \rangle = \frac{1}{2\pi} \int_{-\infty}^{\infty} \int_{-\infty}^{\infty} \langle \hat{a}^\dag(\omega) \hat{a}(\omega') \rangle \mathrm{d}\omega \mathrm{d}\omega'
\end{eqnarray}
we call the integral over the blue-shifted sideband $\langle \delta\hat{a}^\dag \delta\hat{a} \rangle_+$, and the integral over the red-shifted sideband  $\langle \delta\hat{a}^\dag \delta\hat{a} \rangle_-$. The difference between the two sidebands, written as a rate of photons leaving the cavity is thus

\begin{equation}
\begin{split}
&2\kappa\left( \bra \delta\adag \delta\ahat \ket_- - \bra \delta\adag \delta\ahat \ket_+ \right) =\\
&\frac{2\kappa}{2\pi} \int_{-\infty}^{\infty} d\omega \frac{\lambda_\textrm{eff}^2}{\tilde{\Delta}_{c,\textrm{eff}}^2+\kappa^2}\frac{\omega_0}{\omega_s} \frac{1}{(\omega_s+\omega)^2+\gamma^2}\left( 2\gamma (\bar{n}_T+1) + \frac{2\kappa \lambda_\textrm{eff}^2}{\tilde{\Delta}_{c,\textrm{eff}}^2+\kappa^2}\frac{\omega_0}{\omega_s} \right) - \\
&-\frac{2\kappa}{2\pi} \int_{-\infty}^{\infty} d\omega \frac{\lambda_\textrm{eff}^2}{\tilde{\Delta}_{c,\textrm{eff}}^2+\kappa^2}\frac{\omega_0}{\omega_s} \frac{1}{(\omega_s-\omega)^2+\gamma^2}\left( 2\gamma \bar{n}_T + \frac{2\kappa \lambda_\textrm{eff}^2}{\tilde{\Delta}_{c,\textrm{eff}}^2+\kappa^2}\frac{\omega_0}{\omega_s} \right)\\
&=\frac{2\kappa}{2\pi} \int_{-\infty}^{\infty} d\omega \frac{\lambda_\textrm{eff}^2}{\tilde{\Delta}_{c,\textrm{eff}}^2+\kappa^2}\frac{\omega_0}{\omega_s} \frac{2\gamma}{(\omega_s+\omega)^2+\gamma^2} \\
&= 2\kappa \frac{\omega_0}{\omega_s}\frac{\lambda_\textrm{eff}^2}{\tilde{\Delta}_{c,\textrm{eff}}^2+\kappa^2}\label{eq:sidebands}
\end{split}
\end{equation}

Comparing this result with equation (\ref{eq:cdagc}), we find a direct relation between the sideband asymmetry and the expectation value for the number of quasi-particles:

\begin{equation}\label{eq:number_of_QP}
2\kappa\left( \bra \delta\adag \delta\ahat \ket_- - \bra \delta\adag \delta\ahat \ket_+ \right)  = 2\gamma\left( \bra \cdag \chat \ket - \bar{n}_T\right)\,.
\end{equation}

This equation can be interpreted as a rate equation: quasi-particles can be created either via thermal excitation at rate $2\gamma \bar{n}_T$  or via the loss of a red-shifted photon out of the cavity at rate $2\kappa \bra \delta\adag \delta\ahat \ket_-$. On the other hand, quasi-particles can be annihilated via the loss of a  blue-shifted cavity-photon at rate $2\kappa \bra \delta\adag \delta\ahat \ket_+$, or via damping of the $c$-mode at rate $2\gamma\bra \cdag \chat \ket$. These different processes are illustrated in Supplementary Figure \ref{fig:levelscheme} together with an energy diagram of the coupled system.

\subsection{Supplementary Note 2: Data evaluation}
\subsubsection{The dynamic structure factor}
For the Fourier analysis, the detected signal is cut into subtraces of length $\unit[0.02] {P/P_{\mathrm{cr}}}$ corresponding to time windows of $\unit[11]{ms}$. For averaging purposes, each subtrace overlaps halfway with the neighboring ones. The two digitized heterodyne quadratures $Q_1$ and $Q_2$ are first electronically demodulated at $\unit[50]{kHz}$. For each subtrace, we compute the Fourier transform of $Q_1 + i Q_2$ using
frequency bins of $\unit[90]{Hz}$. We normalize the Fourier transforms to the number of data points in the subtrace and the frequency resolution, and average over all 147 experimental runs. The obtained spectral density $\mathcal{S}(\omega)$ of the intracavity photons is converted into a power spectral density of the light field leaking out of the cavity via $PSD = 10\log{(\mathcal{S}(\omega)\cdot h c/\lambda_{\mathrm{c}} \cdot 2 \kappa/1\mathrm{mW})}$ with the cavity decay rate $\unit[\kappa = 2 \pi \cdot 1.25]{MHz}$ and the energy per cavity photon of $h c/\lambda_{\mathrm{c}}$.

\subsubsection{Sideband fitting procedure}
A symmetric low-frequency noise feature develops around the strong coherent field in the self-organized phase, which we attribute to technical phase noise between the signal and the LO beam. Sufficiently deep in the self-organized phase, its amplitude is proportional to the coherent field intensity. 
Before fitting, the scaled averaged technical noise is subtracted from the data in a region up to $\pm \unit[2]{kHz}$. 
Then, the sidebands appearing in the dynamic structure factor in Fig.$~$2 are fitted with two damped Lorentzian functions according to Eq. \ref{eq:spectrum}: $S_{incoh}(\omega)=
BG + \frac{A_{+}}{((\omega+\omega_s)^2+\gamma^2)^2} + \frac{A_{-}}{((\omega-\omega_s)^2+\gamma^2)^2}$, where BG is the heterodyne noise level with the signal path blocked, and $A_{+/-}$ the sideband amplitudes.

The imperfect modeling of the technical noise around the phase transition 
does not permit to start the fit of the sidebands at zero frequency.
In order to be dominated by the sideband amplitude and not by noise, we thus restrict the sideband
fitting routine to a finite frequency $\nu_{\mathrm{min,fit}}$.
This introduces a minimum excitation frequency that can be obtained
through the fit. This effect can be observed in Fig.$~$4 for the quasi-particle frequency very close to the
critical point: it is limited to 1 kHz. 
To account for the influence of the cut-off frequency $\nu_{\mathrm{min,fit}}$, 
we extract the fit results for $\nu_{\mathrm{min,fit}}$ from 810 Hz (900 Hz in the superradiant phase) to
1170 Hz in 90 Hz steps. Their mean is used as the output of the fitting routine.
To determine the error on the fit parameters we select the maximum of one of the three following error estimates: the mean individual fit error, the standard deviation of the fit results for the set of $\nu_{\mathrm{min,fit}}$, or the standard deviation of the fit results for keeping versus not keeping the quasi-particle frequency fixed to the value of the \textit {ab-initio} theory.

\subsection{Supplementary Note 3: Comparison of \textit {ab-initio} theory with experimental data}
The \textit {ab-initio} calculations used in Fig.$~$1-5 are based on Hamiltonian Eq. (\ref{eq:man-body-H}) with no free fit parameter \cite{Mottl2012a}. 
They include the two degenerate cavity TEM$_{00}$ modes with circular polarizations $\epsilon_1$ and $\epsilon_2$ that are driven off-resonantly by the transverse laser beam, which is linearly polarized along the $y$-axis. For the atoms, prepared in the hyperfine state $(F, m_F) = (1, -1)$ with respect to a quantization axis pointing along the cavity
axis (x-axis), the ratio of the corresponding two-photon Rabi frequencies is $\eta_{1}/ \eta_{2} = 2.66/ 0.82$. Here, we denoted $F$ as the total angular momentum and $m_F$ the magnetic quantum number. A single maximally coupled atom induces a maximum dispersive shift of the two cavity modes of $U_0^{1} =
2\pi\times \unit[71]{Hz}$ and $U_0^{2} = 2\pi\times \unit[22]{Hz}$. 

We furthermore take into account the measured coherent intracavity photon number and the transverse pump potential that creates a lattice depth of $2.8 E_\mathrm{r}$ at the critical point, calibrated using Raman-Nath diffraction \cite{Morsch2006}. Systematic uncertainties of these quantities are estimated to be $5\%$ each (see shaded regions in Fig.$~$4 and 5).
The Gaussian envelopes of the pump and cavity fields along the transverse directions \cite{Baumann2010}
are effectively included by weighted averages of $V_\mathrm{p}$, $\eta_{1,2}$ and $U_0^{1,2}$ over the spatial extent
of the atomic cloud. Here, the Thomas-Fermi radii of the condensate in the external harmonic trapping potential is calculated in the presence of the changing transverse pump lattice and atom number, and is given by $(R_x, R_y,R_z) = (3.5,8.3,6.9)$ \textmu m at the critical pump strength, assuming an atom number of $N =1.0 \times 10^5$.

We use our theoretical model to calculate the dynamic structure factor in Fig.$~$2 from the PSD using Eq. (1) in the main text with a conversion error of $5\%$, and to calculate the energy $\omega_0$ of the uncoupled system. The relative coupling strength $\lambda/\lambda_{\mathrm{cr}}$ shown in the inset of Fig.$~$3 is calculated from the experimentally determined quasi-particle frequency $\omega_{s}$ and the theoretically obtained frequency of the uncoupled system, $\omega_0$, using $\lambda^2/\lambda_\mathrm{cr}^2 = 1 - \omega_s^2/ \omega_0^2$. The energy of the quasi-particle mode shown in Fig.$~$4 is evaluated using Eq. (\ref{eq:omega_s}). 
The theoretical number of quasi-particles in Fig.$~$5 is calculated using Eq.$~$(\ref{eq:sidebands}) and (\ref{eq:number_of_QP}). The thermal occupancy of the quasi-particle mode shown is evaluated using $\bar{n}_T = \left(\exp[{\frac{\hbar \omega_s}{k_B T}}]-1\right)^{-1}$, where $k_B$ is the Boltzmann constant, and $T = \unit[38]{nK}$.
The atomic damping rate $\gamma$ is fitted with a phenomenological function based on a dispersion function: $\gamma (x) = p_0(x-1)/(p_1^2 + (x-1)^2)^{p_2} $. The fit results in the normal phase are $(p_0, p_1, p_2) = (\unit[-2\pi \cdot 33.8]{Hz}, 0.039, 1.02)$ and in the self-organized phase, $(p_0, p_1, p_2) = (\unit[2\pi \cdot 299]{Hz}, 0.016, 0.649)$. This function is used in Fig.$~$5 to evaluate the theoretical expectation for the quasi-particle number (grey shaded area).

\end{document}